\documentstyle[floats,prd,aps,eqsecnum,axodraw,12pt]{revtex}
\makeatletter \newbox\tempboxa
\newdimen\captionboxsubcount
\def\capsize#1{\captionboxsubcount=#1pt} \newdimen\captionboxsub
\captionboxsub=\hsize \advance\captionboxsub by -\captionboxsubcount
\advance\captionboxsub by -\captionboxsubcount \long \def\@makecaption#1#2{
\setbox\@tempboxa\hbox{#1 #2} \ifdim \wd\@tempboxa \captionboxsub
\rightskip=\captionboxsubcount \leftskip=\captionboxsubcount #1 #2 \else
\hbox to\hsize{\hfil\box\@tempboxa\hfil} \fi} \makeatother \capsize{30}

\begin{document}
\begin{titlepage}
\begin{flushright}
\begin{minipage}{5cm}
\begin{flushleft}
\small
\baselineskip = 13pt YCTP-P14-99\\hep-ph/9906555
\end{flushleft}
\end{minipage}
\end{flushright}
\begin{center}
\Large\bf Enhanced Global Symmetries  \\ and The Chiral Phase Transition
\end{center}
\footnotesep = 12pt
\begin{center}
\large Thomas {\sc Appelquist }\footnote{Electronic address: {\tt
thomas.appelquist@yale.edu}} P.~S. {\sc Rodrigues da
Silva}\footnote{ Electronic address: {\tt psr7@pantheon.yale.edu}}
Francesco {\sc Sannino}\footnote{ Electronic address : {\tt
francesco.sannino@yale.edu}}
\\~\\
{\it
Department~of~Physics,~Yale~University,~New~Haven,~CT~06520-8120,~USA.}

\end{center}

\begin{center}
\bf Abstract
\end{center}
\begin{abstract}
\baselineskip = 17pt
We examine the possibility that the physical spectrum of a
vector-like gauge field theory exhibits an enhanced global
symmetry near a chiral phase transition. A transition from the
Goldstone phase to the symmetric phase is expected as the number
of fermions $N_f$ is increased to some critical value. Various
investigations have suggested that a parity-doubled spectrum
develops as the critical value is approached. Using an effective
Lagrangian as a guide, we note that parity doubling is associated
with the appearance of an enhanced global symmetry in the
spectrum of the theory. The enhanced symmetry would develop as
the spectrum splits into two sectors, with the first exhibiting
the usual pattern of a spontaneously broken chiral symmetry, and
the second exhibiting an additional, unbroken symmetry and parity
doubling. The first sector includes the Goldstone bosons and
other states such as massive scalar partners. The second includes
a parity-degenerate vector and axial vector along with other
possible parity partners. We note that if such a near-critical
theory describes symmetry breaking in the electroweak theory, the
additional symmetry suppresses the contribution of the parity
doubled sector to the $S$ parameter.

\end{abstract}
\begin{flushleft}
\end{flushleft}
\vfill
\end{titlepage}

\section{Introduction}

  Gauge field theories exhibit many different patterns of
  infrared behavior. During the past few years, there has
  been much progress in understanding the possibilities
   in supersymmetric gauge theories~\cite{IS}.
For non-supersymmetric gauge theories, less is known, but it is
expected that the infrared behavior will vary according to the
number of massless fermions ($N_f$) coupled to the gauge fields.
For a vector-like theory such as QCD, it is known that for low
values of $N_f$, the theory confines and chiral symmetry breaking
occurs. On the other hand, for large $N_f$ the theory loses
asymptotic freedom. In between, there is a conformal window where
the theory does not confine, chiral symmetry is restored, and the
theory acquires a long range conformal symmetry. It has been
proposed that for an $SU(N)$ gauge theory, there is a transition
from the confining, chirally broken theory to the chirally
symmetric theory at $N_f
\approx 4 N$~\cite{ATW,SS}. Recent lattice simulations, however, seem to
indicate ~\cite{mawhinney} that the amount of chiral symmetry
breaking decreases substantially (for $N=3$) when $N_f$ is only
about $4$.

Assuming that a single transition takes place at some critical
value of $N_f$, we can ask questions about the spectrum of the
theory near the transition. In Reference~{\cite{AS}}, it was argued
by studying Weinberg spectral function sum rules that for
near-critical theories parity partners become more degenerate than
in QCD-like theories. This leads naturally to the idea that parity
doublets might form as chiral symmetry is being restored. Lattice
studies also indicate such a possibility~\cite {mawhinney}.

 In this paper we observe using an effective-Lagrangian as a
guide, that the formation of degenerate parity partners is
associated with the appearance of an enhanced global symmetry in
the spectrum of states. We also note that this new symmetry could
play a key role in describing a possible strong electroweak Higgs
sector. Whether the new symmetry can be shown to emerge
dynamically from an underlying gauge theory with $N_f$ near a
critical value remains an open question.

It is worth noting that there exist examples of extra symmetries,
not manifestly present in the underlying theory, but dynamically
generated at low energies. For instance, by using duality
arguments, it has been argued~\cite{LS} that a supersymmetric
$SU(2)$ gauge theory with $N_f$ matter fields and global symmetry
$SU(2N_f)$ is dual to a $SU(N_f -2)$ gauge theory with $N_f$
matter fields. For $N_f \geq 5$, the ultraviolet flavor symmetry
of the latter theory is $SU_L(N_f)\times SU_R(N_f)\times U_B(1)$.
Since its infrared global symmetry must be $SU(2N_f)$ (that of
the dual), its infrared symmetry is enhanced.

In Section \ref{ELSUSU} we discuss the appearance of enhanced
global symmetry. Confinement is assumed and the symmetry of the
underlying gauge theory, $SU_L(N_f)\times SU_R(N_f)$, is built
into an effective Lagrangian describing the physical states of
the theory. Parity invariance is imposed and the usual pattern of
chiral symmetry breaking ($SU_L(N_f)\times SU_R(N_f)\rightarrow
SU_V(N_f)$) is assumed. The $N_f^2-1$ Goldstone bosons appear
together with scalar chiral partners. We augment the spectrum
with a set of vector fields for both the $SU_L(N_f)$ and
$SU_R(N_f)$ symmetry groups. The Lagrangian thus takes the form
of a linear sigma model coupled to vectors. It could be expanded
to include fields corresponding to other states as well. The
natural mass scale of this strongly interacting system is
expected to be of order $2\pi v$, where $v$ is the vacuum
expectation value.

We examine the spectrum and recognize that there is a particular
choice of the parameters that allows for a degenerate vector and
axial-vector, while enlarging the global symmetry to include an
additional (unbroken) $SU_L(N_f)\times SU_R(N_f)$. This happens
as the spectrum of the theory splits into two sectors with one
displaying the additional symmetry. We then briefly review the
arguments (see Ref.~{\cite{AS}}) that an underlying near-critical
$SU(N)$ gauge theory might naturally lead to a more degenerate
vector-axial spectrum than in QCD, and to an enhanced symmetry.
Finally we note that even a discrete additional symmetry,
$Z_{2L}\times Z_{2R}$, of the effective theory is adequate to
insure the mass degeneracy of the vector and axial
vector\footnote{We thank Noriaki Kitazawa for suggesting the
possibility of a discrete symmetry.}.

The possible appearance of an additional, continuous symmetry was
considered by Casalbuoni et al in Refs.{\cite{sei,sei2}}. These
papers were restricted to the case $N_f = 2$ and did not include
discussion of the possible connection to a near-critical underlying
theory. The treatment in Ref.~\cite{sei} made use of a nonlinear
realization for the Goldstone degrees of freedom, using hidden
gauge symmetry methods~\cite{BKY}. We could generate the effective
Lagrangian of Ref.~\cite{sei} by integrating out the massive scalar
degrees of freedom, but that would keep some massive degrees of
freedom (the vector fields) and neglect others. When we focus on
low energy consequences (in Section \ref{ELWSU}), we will integrate
out all the massive degrees of freedom leading to the electroweak
chiral Lagrangian. The treatment of Ref.~\cite{sei2} utilized a
linear realization for the scalars and focused on the decoupling of
the vectors as they are made heavy relative to the weak scale. We
do not take this limit here since we assume the vector and scalar
masses to be of the same order.

In Section \ref{ELWSU} we embed the electroweak gauge group
within the global symmetry group. We observe that the enhanced
symmetry of the strongly interacting sector, which now provides
electroweak symmetry breaking, plays an important role. The
additional symmetry is a partial custodial symmetry for the
electroweak $S$ parameter, in the sense that the parity doubled
part of the strong sector, by itself, makes no contribution to
$S$. This is shown by integrating out the massive physics to
construct the terms in the low energy electroweak chiral
Lagrangian. The $S$ parameter corresponds to one such term.

We extend the study to fermions in a pseudoreal representation of
the underlying gauge group in Section \ref{ELSP}. In this case
parity is automatically enforced. The pseudoreal representations
allow for the lowest number of colors (i.e. $N=2$) and consequently
for the lowest possible number of flavors for which the theory
might show a dynamically enhanced symmetry. The enhanced global
symmetry is $\left[ SU(2N_f)\right]^2$ spontaneously broken to
$Sp(2N_f)\times SU(2N_f)$.

In Section \ref{conc} we conclude and suggest some directions for
future work. In Appendix \ref{agenerators} we provide an explicit
representation for the $Sp(4)$ generators.

\section{Effective Lagrangian for $SU_L(N_f)\times SU_R(N_f)$ global symmetry}

\label{ELSUSU}

To discuss the possible appearance of enhanced symmetry in a
strongly interacting spectrum, some description of the spectrum
is needed. We will find it helpful to use an effective Lagrangian
possessing $SU_L(N_f)\times SU_R(N_f)$ symmetry, the global
invariance of the underlying gauge theory. We assume that chiral
symmetry is broken according to the standard pattern $
SU_L(N_f)\times SU_R(N_f)
\rightarrow SU_V(N_f)$. The $N_f^2
-1$ Goldstone bosons are encoded in the $N_f\times N_f$ real traceless
matrix $\Phi^i_j$ with $i,j=1,..., N_f$ . The complex matrix
$M=S+i\Phi$ describes both the Goldstone bosons as well as
associated scalar partners $S$. It transforms linearly under a
chiral rotation:
\begin{equation}
M\rightarrow u_L M u^{\dagger}_R\;,
\end{equation}
with $u_{L/R}$ in $SU_{L/R}(N_f)$.

To augment the massive spectrum, we introduce vector and axial
vector fields following a method outlined in Ref.~\cite{joe}. We
first formally gauge the global chiral group introducing the
covariant derivative
\begin{equation} D^{\mu} M =
\partial^{\mu} M - i\tilde{g} A^{\mu}_{L} M + i \tilde{g} M A^{\mu}_R \ ,
\end{equation}
where $A_{L/R}^{\mu}=A_{L/R}^{\mu,a}T^a$ and $T^a$ are the
generators of $SU(N_f)$, with $a=1,...,N_f^2-1$ and $
\displaystyle{{\rm Tr}\left[T^a T^b\right]=\frac{1}{2} \delta^{ab}}$. The
left and right couplings are the same since we assume parity
invariance. Under a chiral transformation
 \begin{equation}
  A_{L/R}^{\mu} = u_{L/R} A_L^{\mu} u_{L/R}^{\dagger} -
\frac{i}{\tilde{g}}
\partial^{\mu}u_{L/R} u^{\dagger}_{L/R}.
\label{gaugetrans}
\end{equation}

The effective Lagrangian needs only to be invariant under global
chiral transformations. Including terms only up to mass dimension
four, it may be written in the form \begin{eqnarray}
L=&~&\frac{1}{2}\,{\rm Tr}
\left[D_{\mu}M D^{\mu}M^{\dagger}\right] + m^2 \, {\rm
Tr}\left[A_{L\mu}A_L^{\mu} +A_{R\mu}A_R^{\mu}\right] \nonumber \\
&+& h\, {\rm Tr}\left[A_{L \mu}M A_R^{\mu}M^{\dagger}\right]
+{r}\,{\rm Tr}
\left[A_{L\mu}A_L^{\mu}MM^{\dagger} + A_{R\mu}A_R^{\mu}M^{\dagger}M
\right] \nonumber \\ &+& i\, \frac{s}{2}\,{\rm Tr}\left[A_{L\mu}\left(M
D^{\mu}M^{\dagger}-D^{\mu}M M^{\dagger} \right) +
A_{R\mu}\left(M^{\dagger}D^{\mu}M -
D^{\mu}M^{\dagger}M\right)\right] \ .
\label{asLagrangian}
\end{eqnarray}
The parameters $h,r$ and $s$ are dimensionless real parameters,
while $m^2$ is a common mass term. To this, we may add a kinetic
term for the vector fields
\begin{equation}
L_{{\rm Kin}}=-\frac{1}{2}\,{\rm Tr}\left[F_{L\mu \nu} F^{\mu
\nu}_L +
F_{R\mu \nu} F^{\mu \nu}_R \right] \ ,  \label{kin}
\end{equation} where
\begin{equation} F^{\mu \nu}_{L/R}=
\partial^{\mu}A_{{L/R}}^{\nu}-\partial^{\nu} A_{{L/R} }^{\mu} - i
\tilde{g}
\left[A_{{L/R}}^{\mu},A_{{L/R}}^{\nu}\right]\ , \label{F} \end{equation}
along with vector-interaction terms respecting only the global
symmetry. Finally, we may add the double trace term,
\begin{equation}
\displaystyle{{\rm Tr}\left[M M^{\dagger}\right]{\rm Tr}\left[A_L^2 + A_R^2
\right]}\ ,  \label{tracedouble}
\end{equation}
at the dimension-four level. To arrange for symmetry breaking, a
potential $V(M,M^{\dagger})$ must be added. When the effective
Lagrangian is extended to the dimension-six level and higher,
many new operators enter.

Parity is also a symmetry and it acts on the fields according to
\begin{eqnarray} P\, M(\mbox{\boldmath${x}$}) \, (P)^{-1} &=&
M^{\dagger}(- \mbox{\boldmath${x}$}) \ , \\ P\,
A_{L/R}^{\mu}(\mbox{\boldmath${x}$}) \, (P)^{-1} &=&
\epsilon({\mu}
)A_{R/L}^{\mu} (-\mbox{\boldmath${x}$}) \ , \end{eqnarray} where
$\epsilon({\mu})=1$ for ${\mu}=0$ and $-1$ for ${\mu=1,2,3}$.

The spectrum described by this effective Lagrangian consists of
Goldstone bosons, a set of scalars, and massive vector and axial
vectors.  With its massive vectors and axial vectors, it is of
course not renormalizable, but it can nevertheless provide a
reasonable description of low-lying states. (It is worth noting
that a Lagrangian of this type does this for the low-lying QCD
resonances~\cite {effective Lagrangians}). While it cannot be a
complete description of the hadronic spectrum, it has sufficient
content to guide a general discussion of enhanced symmetries.

Keeping only terms quadratic in the fields and temporarily
neglecting the massive scalars, the Lagrangian Eq.
(\ref{asLagrangian}) takes the form
\begin{eqnarray} L=&~&\frac{1}{2} {\rm Tr}\left[\partial_{\mu}\Phi
\partial^{\mu}\Phi\right] + \sqrt{2}(s-\tilde{g}) \,{v}\, {\rm
Tr}\left[\partial_{\mu}\Phi A^{\mu}
\right] + M^2_A {\rm Tr}\left[A_{\mu}A^{\mu}\right] + M^2_V {\rm Tr}\left[
V_{\mu}V^{\mu}\right] ,
  \label{quadratic}
\end{eqnarray}
where $M= v + i\Phi $, $v$ is the vacuum expectation value and we
have defined the new vector fields
\begin{equation}
V=\frac{A_L + A_R}{\sqrt{2}}\ , \qquad\qquad A=\frac{A_L -
A_R}{\sqrt{2}}
\ .
\label{va}
\end{equation}
The vector and axial masses are related to the effective
Lagrangian parameters via
\begin{eqnarray}
M^2_A&=& m^2 + v^2\,\left[r + \tilde{g}^2 - 2\,s\,\tilde{g} -
\frac{h}{2}
\right] \ ,  \nonumber \\ M^2_V &=& m^2 + v^2 \left[r + \frac{h}{2}
\right] \,
\end{eqnarray}
where the contribution from Eq.~(\ref{tracedouble}) has been
absorbed into $m^2$. The terms proportional to $v^2$ are
Higgs-like contributions, arising from the spontaneous breaking.

The second term in Eq.~(\ref{quadratic}) mixes the axial vector
with the Goldstone bosons. This kinetic mixing may be
diagonalized away by the field redefinition
\begin{equation}
A\rightarrow A + v\frac{\tilde{g}-s}{\sqrt{2}M^2_A} \partial \Phi
\ ,
\end{equation} leaving the mass spectrum unchanged~\cite{joe}.
The vector-axial vector mass difference is given by
\begin{equation}
M^2_A - M^2_V = v^2\, \left[\tilde{g}^2 - 2 \tilde{g}\,s - h
\right] \ .
\label{differenza}
\end{equation}
In QCD this difference is known experimentally to be positive, a
fact that can be understood by examining the Weinberg spectral
function sum rules (see Ref.~{\cite{BDLW}} and references
therein). The effective Lagrangian description is of course
unrestrictive. Depending on the values of the $\tilde{g}$, $s$
and $h$ parameters, one can have a degenerate or even inverted
mass spectrum.

What kind of underlying gauge theory might provide a degenerate
or inverted spectrum? Clearly, it has to be different from QCD,
allowing for a modification of the spectral function sum rules.
In Reference~\cite{AS}, an $SU(N)$ gauge theory (with $N>2$) and
$N_f$ flavors was considered.  If $N_f$ is large enough but below
$11N/2$, an infrared fixed point of the gauge coupling $\alpha_*$
exists, determined by the first two terms in the $\beta$
function. For $N_f$ near $11N/2$, $\alpha_*$ is small and the
global symmetry group remains unbroken. For small $N_f$, on the
other hand, the chiral symmetry group $
\displaystyle{SU_L(N_f)\otimes SU_R(N_f)}$ breaks to its diagonal subgroup.
One possibility is that the transition out of the broken phase
takes place at a relatively large value of $N_f/N (\approx 4)$,
corresponding to a relatively weak infrared fixed
point~\cite{ATW,SS}. An alternate possibility is that the
transition takes place in the strong coupling regime,
corresponding to a small value of $N_f/N$~\cite {mawhinney}. The
larger value emerges from the renormalization group improved gap
equation, as well as from instanton effects~\cite{ASe}, and
saturates a recently conjectured upper limit~\cite{ACS}. It
corresponds to the perturbative infrared fixed point $\alpha_*$
reaching a certain critical value $\alpha_c$. A similar result
has also been obtained by using a suitable effective Lagrangian
\cite{SS}.

These studies also suggest that the order parameter, for example
the Goldstone boson decay constant $F_{\pi}\equiv v$, vanishes
continuously at the transition relative to the intrinsic
renormalization scale $\Lambda$ of the gauge theory. In the broken
phase near the transition, the fact that one is approaching a phase
with long range conformal symmetry suggests that all massive states
scale to zero with the order parameter relative to $\Lambda$
~\cite{Sekhar}.

In Reference~{\cite{AS}} the spectrum of states in the broken phase
near a large-$N_f/N$ transition was investigated using the spectral
function sum rules. It was shown that the ordering pattern for
vector-axial hadronic states need not be the same as in QCD-like
theories (small $N_f/N$). The crucial ingredient is that these
theories contain an extended "conformal region" extending from
roughly $2\pi F_{\pi}$ to the scale $\Lambda$ where asymptotic
freedom sets in. In this region, the coupling remains close to an
approximate infrared fixed point and the theory has an approximate
long range conformal symmetry. It was argued that this leads to a
reduced vector-axial mass splitting, compared to QCD-like theories.
This suggests the interesting possibility that parity doublets
begin to form as chiral symmetry is being restored. That is, the
vector-axial mass ratio approaches unity as the masses decrease
relative to $\Lambda$. Lattice results seem to provide supporting
evidence for such a possibility~\cite{mawhinney}, although at
smaller values of $N_f/N$.

If a parity doubled spectrum does appear, it is natural to expect
it to be associated with some new global symmetry. While we have
not demonstrated the appearance of a new global symmetry using
the underlying degrees of freedom, we can explore aspects of
parity doubling at the effective Lagrangian level. Returning to
this description, we note that vector-axial parity doubling
corresponds to the parameter choice (see Eq. (\ref{differenza})),
\begin{equation}
\tilde{g}^2 = 2 \tilde{g}\,s + h \ .
\label{degenerate}
\end{equation}
This condition does not yet reveal an additional symmetry and
therefore there is no reason to expect parity degeneracy to be
stable in the presence of quantum corrections and the many higher
dimensional operators that can be added to the effective
Lagrangian in Eq.~(\ref{asLagrangian}).

However, for the special choice $s = \tilde{g}$, $r
=\tilde{g}^2/2$
and $h=-\tilde{g}^2$, the effective Lagrangian acquires a new
continuous global symmetry that protects the vector-axial mass
difference. The effective Lagrangian at the dimension-four level
takes the simple form
\begin{equation} L=\frac{1}{2}\,{\rm Tr}
\left[\partial_{\mu}M \partial^{\mu}M^{\dagger} \right] + m^2 \, {\rm
Tr}\left[A_{L\mu}A_L^{\mu} +A_{R\mu}A_R^{\mu}\right],
  \label{dlag}
\end{equation}
along with vector kinetic and interaction terms, the interaction
term Eq.~(\ref{tracedouble}), and the symmetry breaking potential
$V(M,M^{\dagger})$. The theory now has two sectors, with the vector
and axial vector having their own unbroken global $SU_L(N_f)\times
SU_R(N_f)$. The two sectors interact only through the product of
singlet operators. The full global symmetry is
$\left[SU_L(N_f)\times SU_R(N_f)\right]^2\times U_V(1)$
spontaneously broken to $ SU_V(N_f)\times U_V(1) \times
\left[SU_L(N_f)\times SU_R(N_f)\right]$. The vector and axial
vector become stable due to the emergence of a new conservation
law. This enhanced symmetry would become exact only in the chiral
limit. For finite but small (relative to $\Lambda$) values of the
mass scales in Eq. (\ref{dlag}), there are additional, smaller
terms giving smaller mass splittings and small width-to-mass
ratios.

It is of course a simple observation that a new symmetry and
conservation law emerge if a theory is split into two sectors by
setting certain combinations of parameters to zero. But here we
were led to this possibility by looking for a symmetry basis for
the parity doubling that has been hinted at by analyses of the
underlying gauge theory. Although we have used a relatively
simple effective Lagrangian, we anticipate that the conclusion is
true in general, that is, that parity doublets form in the
spectrum of a strongly interacting theory with chiral symmetry
breaking only if the spectrum splits into two sectors, one
exhibiting the spontaneous breaking and the other, parity
doubled, sector exhibiting an unbroken additional symmetry.

We next observe that along with the additional global symmetry
$SU_L(N_f)\times SU_R(N_f)$, the effective Lagrangian Eq.
(\ref{dlag}) possesses a discrete $Z_{2L}\times Z_{2R}$ symmetry.
Under $Z_{2L}\times Z_{2R}$ the vector fields transform according
to
\begin{equation} A_L \rightarrow z_{L} A_L \ , \qquad A_R \rightarrow
z_{R}A_R \ , \end{equation} with $z_{L/R}=1,-1$ and $z_{L/R}\in
Z_{2L/R}$. Actually, the discrete symmetry alone is enough to
insure vector-axial mass degeneracy and stability against decay.
In that case, additional interaction terms, such as the single
trace term
\begin{equation}  r{\rm Tr} \left[A_{\mu L} A^{\mu}_L M M^{\dagger} +
A_{\mu R} A^{\mu}_R M^{\dagger} M \right] \ ,
\end{equation}
are allowed, but degeneracy and stability are still insured.  Of
course, trilinear vector interactions will not respect this
discrete symmetry. Nevertheless, one can not rule out the
possibility that it is only this smaller, discrete symmetry that
appears as an effective infrared symmetry of an underlying gauge
theory near the chiral/conformal transition.

 From the point of view of the underlying theory, the appearance
of any additional symmetry in the spectrum, at criticality, would
seem mysterious. The composite degrees of freedom in both sectors
are made of the same fundamental fermions with a single
underlying $SU_L(N_f)\times SU_R(N_f)$ symmetry. If the symmetry
of the parity doubled sector is an unbroken $SU_L(N_f)\times
SU_R(N_f)$, it would look as though the chiral symmetry is being
realized there in the Wigner-Weyl mode. If that is the case,
chiral dynamics would have to be influenced by confinement and
bound state formation in an interesting new way. Whether a
near-critical gauge theory can lead to this behavior is an
unresolved question.

\section{Strongly Interacting Electroweak Sector} \label{ELWSU}

  We next discuss the consequences of enhanced symmetry for a
strong symmetry breaking sector of the standard electroweak theory,
embedding the $SU_L(2) \times U_Y(1)$ gauge symmetry in the global
$SU_L(N_f)\times SU_R(N_f)$ chiral group. In this Section, for
simplicity, we will restrict attention to the $SU_{L}(2)\times
SU_R(2) $ subgroup of the full global group \cite{sei2}. The
electroweak gauge transformation then takes the form
\begin{equation} M \rightarrow u_W M u_Y^{\dagger} \ ,
\end{equation}
where $M$ is now a $2 \times 2$ matrix which can be written as
$\displaystyle{M=\frac{1}{\sqrt{2}} \left[\sigma +i
\vec{\tau}\cdot\vec{\pi}\right]}$ , where
$\displaystyle{u_W=u_L=\exp\left(\frac{i}{2}\epsilon^a
\tau^a\right)}$
with $\tau^a$ the Pauli matrices, and where
$\displaystyle{u_Y=\exp
\left(
\frac{i}{2}\epsilon_0 \tau^3\right)}$. The weak vector boson fields
transform as
\begin{eqnarray}
W^{\mu}&\rightarrow& u_L W^{\mu}u_L^{\dagger} -
\frac{i}{g}\partial^{\mu}u_L u^{\dagger}_L \ , \\ B^{\mu}&\rightarrow&
u_Y B^{\mu}u_Y^{\dagger} - \frac{i}{g^{\prime}} \partial^{\mu}u_Y
u^{\dagger}_Y \ ,  \label{rotation} \end{eqnarray} where $g$ and
$g^{\prime}$ are the standard electroweak coupling constants,
$\displaystyle{W_{\mu}=W_{\mu}^a \frac{\tau^a}{2}}$ and
$\displaystyle{B_{\mu}=B_{\mu}\frac{\tau^3}{2}}$.

A convenient method of coupling the electroweak gauge fields to
the globally invariant effective Lagrangian of Section II is to
introduce a covariant derivative, which includes the $W$ and $B$
fields as well as the strong vector and axial-vector fields,
\begin{equation} {\rm
D}^{\mu}M=\partial^{\mu}M - ig W^{\mu}M + i g^{\prime}MB^{\mu}
-i\tilde{
g}c C^{\mu}_L M +i \tilde{g}c^{\prime}M C^{\mu}_R \ ,
\label{newD}
\end{equation} where we have defined the new vector fields
\begin{equation} C^{\mu}_L=A_L^{\mu} -\frac{g}{\tilde{g}}W^{\mu}\ ,
\qquad C^{\mu}_R=A_R^{\mu} -\frac{g^{\prime}}{\tilde{g}}B^{\mu} \ ,
\label{cdef}
\end{equation}
and where $c$ and $c^{\prime}$ are arbitrary real constants.
Since the $A_{L/R}^{\mu}$ transform as Eq.~(\ref{gaugetrans}),
the $C_{L/R}^\mu$ transform under the electroweak transformations
as
\begin{equation} C^{\mu}_L \rightarrow u_L C^{\mu}_L u_L^{\dagger} \ ,\qquad
C^{\mu}_R \rightarrow u_Y C^{\mu}_R u_Y^{\dagger}.
\end{equation}
 By requiring invariance under the parity operation exchanging
the labels $L\leftrightarrow R$ we have the extra condition
$c=c^{\prime}$.

The effective Lagrangian is constructed to be invariant under a
local $SU_{L}(2) \times U_{Y}(1)$ as well as $CP$. The $CP$
transformation properties of the fields\footnote{Here we
summarize the $CP$ field transformations:
\begin{eqnarray}
CP\, M(\mbox{\boldmath${x}$}) \, (CP)^{-1} &=& \eta M^*(- \mbox{
\boldmath${x}$}) \ , \\
CP\, A_{L/R\, \mu}(\mbox{\boldmath${x}$}) \, (CP)^{-1}
&=&-A_{L/R}^{\mu} (-
\mbox{\boldmath${x}$}) \ , \\
CP\, W_{\mu}(\mbox{\boldmath${x}$}) \, (CP)^{-1} &=&-W^{\mu} (-
\mbox{\boldmath${x}$}) \ , \\ CP\, B_{\mu}(\mbox{\boldmath${x}$}) \,
(CP)^{-1} &=&-B^{\mu} (- \mbox{\boldmath${x}$}) \ ,
\end{eqnarray}
where $\eta$ is an arbitrary $C$ phase.} insure that the
covariant derivative transforms as $M$, i.e.
\begin{equation}
CP\, {\rm D}_{\mu} M(\mbox{\boldmath$x$}) \, (CP)^{-1}= \eta
\left( {\rm
D}
^{\mu} M (-\mbox{\boldmath$x$}) \right)^{*} \ .
\end{equation}
The effective Lagrangian  is then obtained by replacing in Eq.
(\ref {asLagrangian}) the covariant derivative with the new one
in Eq.~(\ref{newD}). To make the theory electroweak gauge
invariant, one substitutes the  $A_{L/R}$ with the $C_{L/R}$,
giving, through dimension four,
\begin{eqnarray}
L=&~&\frac{1}{2}{\rm Tr}\left[ {\rm D}_{\mu }M{\rm D}^{\mu
}M^{\dagger }
\right]+m^{2}\,{\rm Tr}\left[ C_{L\mu }C_{L}^{\mu }+C_{R\mu }C_{R}^{\mu }
\right]  \nonumber \\
&+&h\,{\rm Tr}\left[ C_{L\mu }MC_{R}^{\mu }M^{\dagger
}\right]+{r}\,{\rm Tr}
\left[ C_{L\mu }C_{L}^{\mu }MM^{\dagger }+C_{R\mu }C_{R}^{\mu }M^{\dagger }M
\right]  \nonumber \\
&+&i\,\frac{s}{2}\,{\rm Tr}\left[ C_{L\mu }\left( M{\rm D}^{\mu
}M^{\dagger }-{\rm D}^{\mu }MM^{\dagger }\right) +C_{R\mu }\left(
M^{\dagger }{\rm D}
^{\mu }M-{\rm D}^{\mu }M^{\dagger }M\right) \right] .
\label{newLagrangian}
\end{eqnarray}
To this we add a kinetic term
\begin{equation}
L_{{\rm Kin}}= -\frac{1}{2}{\rm Tr}\left[ F_{L\mu \nu }F_{L}^{\mu
\nu
}+F_{R\mu \nu }F_{R}^{\mu \nu }\right] -\frac{1}{2}{\rm Tr}\left[
W_{\mu
\nu }W^{\mu \nu } \right] -\frac{1}{2}{\rm Tr}\left[ B_{\mu \nu }B^{\mu
\nu }
\right] \ ,
\label{generalkinetic}
\end{equation}
where
\begin{eqnarray}
W_{\mu \nu } &=&\partial _{\mu }W_{\nu }-\partial _{\nu }W_{\mu
}-ig\left[ W_{\mu },W_{\nu }\right] \ ,  \nonumber \\ B_{\mu \nu
} &=&\partial _{\mu }B_{\nu }-\partial _{\nu }B_{\mu }\ ,
\label{WB}
\end{eqnarray}
with the $F_{L/R}$ for the fields $A_{L/R}$ defined in Eq.
({\ref{F}}), along with other interaction terms involving the
$C_{L/R}$ fields, the interaction term
\begin{equation}
\displaystyle{{\rm Tr}\left[M M^{\dagger}\right]{\rm Tr}\left[C_L^2 + C_R^2
\right]},
  \label{Ctracedouble}
\end{equation}
and a symmetry breaking potential.

One can show that this is the most general dimension-four,
CP-invariant Lagrangian describing a strongly interacting set of
scalars, vectors, and axial vectors with a spontaneously broken
$SU_L(2)\times SU_R(2)$ symmetry, and possessing electroweak
gauge invariance. It describes weak mixing between the $A_{L/R}$
fields and the $W$ and $Z$, and, through the mixing, conventional
electroweak charges for the $A_{L/R}$. The extension of this
effective Lagrangian to the relevant case of the larger symmetry
group $SU_L(N_f)\times SU_R(N_f)$ with $N_f > 2$, is
straightforward.

Replacing $M$ by its vacuum value $v/\sqrt{2}$, and keeping only
terms quadratic in the fields, the Lagrangian Eq.
(\ref{newLagrangian}) takes the form
\begin{eqnarray}
L=~ &&M_{A}^{2}{\rm Tr}\left[ A^{2}\right] +M_{V}^{2}{\rm
Tr}\left[ V^{2}
\right] -\frac{\sqrt{2}}{\tilde{g}}(1-\chi )M_{A}^{2}{\rm Tr}
\left[ \left( gW-g^{\prime }B\right) A\right]  \nonumber \\
&-&\frac{\sqrt{2}}{\tilde{g}}M_{V}^{2} {\rm Tr}\left[ \left(
gW+g^{\prime }B\right) V\right]
+\frac{M_{V}^{2}}{2\tilde{g}^{2}}{\rm Tr}
\left[ \left( gW+g^{\prime }B\right) ^{2}\right]  \nonumber \\
&+&\frac{M_{A}^{2}}{2\tilde{g}^{2}}
\left( 1+\delta \right) {\rm Tr}\left[ \left( gW-g^{\prime }B\right)
^{2}\right] +\cdots \ ,
  \label{QLagrangian}
   \end{eqnarray} where we have defined:
\begin{eqnarray}
M_{V}^{2} &=&m^{2}+v^{2}\left[ r+\frac{h}{2}\right] \ ,
\nonumber \\
M_{A}^{2} &=&m^{2}+v^{2}\left[
r+\tilde{g}^{2}c^{2}-2s\,\tilde{g}c-\frac{h}{2}\right] \ ,
\nonumber \\
\chi &=&\frac{v^{2}}{2 M_{A}^{2}}\tilde{g}\left[ \tilde{g}c-s\right] \ ,
\nonumber \\
\delta &=&\frac{v^{2}}{2 M_{A}^{2}}\left[ \tilde{g}^{2}(1-2c)+2s\,\tilde{g}
\right] \ .  \label{elenco}
\end{eqnarray}
The vector $V$ and axial $A$ fields are defined in
Eq.~(\ref{va}). This quadratic Lagrangian describes masses for
the  V and A, weak mass mixing with the $W$ and $B$, and a mass
matrix for the $W$ and $B$. There is no further, kinetic energy
mixing among these fields. The vector and axial vector masses,
$M_V^2$ and $M_A^2$, are arbitrary, depending on the choice of
parameters, although generically we expect them and the scalar
masses to be of order $4\pi^{2}v^2$.

The weak mixing terms in Eq.~(\ref{QLagrangian}) provide a
contribution from physics beyond the standard model to the
oblique electroweak corrections. These may be described by the
$S$, $T$, and $U$ parameters, but the last two vanish in the
present model because there is no breaking of weak isospin in the
strong sector. While this is not apparent in Eq.
(\ref{QLagrangian}), it is insured by the Ward identities and
easily revealed through the mixing effects. The $S$ parameter
receives contributions from all the physics beyond the standard
model, including, in the model being used here, loops of
pseudo-Goldstone bosons (PGB's), the strongly interacting massive
scalars, and the vector and axial vector. The direct,
vector-dominance contribution of the vector and axial vector may
be read off from Eq.~(\ref{QLagrangian}) together with the
kinetic term for the $V$ and $A$. One finds
\begin{equation} S_{vect-dom} = \frac{8\pi}{\tilde{g}^{2}}\left[
\frac{M_{A}^{2}
\left( 1-\chi \right) ^{2}}{M_{Z}^{2}-M_{A}^{2}}-
\frac{M_{V}^{2}}{M_{Z}^{2}-M_{V}^{2}}\right]
 \approx \frac{8\pi}{\tilde{g}^{2}}  \left[1 - \left(1 -
 \chi\right)^2\right] .
\label{s-ciao}
\end{equation}
Clearly, this contribution to the $S$ parameter can take on any
value depending on the choice of parameters. Its typical order of
magnitude, with the strong coupling estimate $\tilde{g}^2 \approx 4
\pi^2$, is expected to be $O(1)$. This expression can be seen to be
equivalent to the familiar vector-dominance formula
$\displaystyle{S_{vect-dom} \approx 4\pi
\left[\frac{F^2_V}{M^2_V}-\frac{F^2_A}{M^2_A}\right]}$
\cite{PT}, with the identifications
$\displaystyle{F^2_V=\frac{2}{\tilde{g}^2} M^2_V}$ and
$\displaystyle{F^2_A=\frac{2}{\tilde{g}^2} M^2_A
\left(1-\chi\right)^2}$.

 We next observe that the choice
\begin{equation} s=\tilde{g}c\ ,\quad
\quad h=-\tilde{g}^{2}c^{2}\,
\end{equation}
gives $\chi = 0$, leading immediately to the degeneracy of the
vector and axial vector (see Eq.~(\ref{elenco})), the relation $F_A
= F_V$, and the vanishing of $S_{vect-dom}$. The further choice
$\displaystyle{r=\frac{\tilde{g}^2c^2}{2}}$ leads to the collapse
of the general effective Lagrangian into the simple form
\begin{equation} L=\frac{1}{2}{\rm Tr}\left[{D}_{\mu }M{D}^{\mu
}M^{\dagger}\right] +m^{2}\, {\rm Tr}\left[ C_{L\mu }C_{L}^{\mu
}+C_{R\mu }C_{R}^{\mu }\right] \ , \label{symmetric}
\end{equation}
along with the kinetic terms of Eq.(\ref{generalkinetic}),
interactions among the $C_{L/R}^\mu$ fields, the interaction term
Eq.(\ref{Ctracedouble}), and a symmetry breaking potential. Here,
$\displaystyle{DM=\partial M-igWM+ig^{\prime}MB}$ is the standard
electroweak covariant derivative, and $C_{L/R}^\mu$ are given by
Eq.~(\ref{cdef}).

The strongly interacting sector has split into two subsectors,
communicating only through the electroweak interactions. One
subsector consists of the Goldstone bosons together with their
massive scalar partners. The other consists of the degenerate
vector and axial vector described by the $A_{L/R}^{\mu }$ fields.
The mass mixing in Eq.~(\ref{newLagrangian}) insures that they
have conventional electroweak couplings. In the absence of
electroweak interactions, there is an enhanced symmetry $\left[
SU_{L}(2)\times SU_{R}(2)\right] \times
\left[ SU_{L}(2)\times SU_{R}(2)\right]$, breaking spontaneously
to $ SU_{V}(2)\times \left[ SU_{L}(2)\times SU_{R}(2)\right]$. The
electroweak interactions explicitly break the enhanced symmetry to
$SU_{L}(2) \times U_Y(1)$. All of this may be generalized to $N_f >
2$, necessary to yield a near-critical theory.

The additional symmetry has an important effect on the $S$
parameter, suppressing contributions that are typically large in
QCD-like theories. It doesn't suppress all contributions, of
course, since the symmetry breaking subsector gives contributions
that are expected to be of order unity. The parity-doubled
subsector, however, cannot by itself contribute to $S$, because
$S$ relies on electroweak symmetry breaking for its existence. It
is the coefficient of an operator in the low-energy electroweak
chiral Lagrangian ($L_1$ in Ref.~\cite{long}), which may be
written in the form ${\rm Tr}~W^{\mu\nu}UB_{\mu\nu}U^{\dagger}$,
where $W^{\mu\nu}$ and $B_{\mu\nu}$ are defined in Eq.~(\ref{WB})
and $U$ is the Goldstone matrix field satisfying the nonlinear
constraint $U U^{\dagger} = U^{\dagger}U = 1$. Clearly the $U$
operator, with its vacuum value $U = 1$, is necessary to couple
$W^{\mu\nu}$ to $B_{\mu\nu}$.

Among the contributions to $S$ remaining in the limit of enhanced
symmetry, are loops of pseudo-Goldstone bosons, present when $N_f
> 2$. They may be estimated using chiral perturbation theory,
with the standard-model corrections removed by convention. While
they typically give contributions to $S$ of order unity, their
specific value depends on details such as mass estimates for the
PGB's that arise from electroweak, QCD, and others
interactions~\cite{PGBandS}. An interesting new feature in the
limit of enhanced symmetry is that the PGB contribution is not
related to a direct, vector-dominance effect (which is now zero).
There will also be contributions from the strongly interacting
TeV physics, represented in our effective Lagrangian by the
massive scalars. Our purpose here is not to make these estimates,
but only to point out that an enhanced symmetry, leading to
vector-axial vector degeneracy, will suppress contributions to
$S$ purely from the parity doubled sector. These include the
typically large vector dominance contribution discussed above.

Finally we note that, as we discussed at the end of Section III,
it could be that only a lesser, discrete symmetry emerges in the
physical spectrum. Even this would be sufficient to insure
vector-axial degeneracy and the vanishing of the vector dominance
contribution to the $S$ parameter. The discrete symmetry of
Section III would only be possible if trilinear vector
interactions are somehow suppressed. It will be interesting to
explore the phenomenology of this possibility, in particular the
effect on the self interactions of the $W$ and $Z$.

\section{$SU(2N_f)$ global symmetry}

\label{ELSP}

In this section we adapt the above discussion to the interesting
case of fermions in pseudoreal representations of the gauge
group. The simplest example is provided by an underlying $SU(2)$
gauge theory, a choice that will also offer the smallest value
for the critical $N_f$~\cite{postmodern}. Such theories are
currently being
 investigated on the lattice (see Ref.~\cite{Mariapaola}). The
quantum global symmetry for $N_f$ matter fields in the pseudoreal
representation of the gauge group~\cite{Peskin} is $SU(2N_f)$. We
expect the gauge dynamics to create a non vanishing
fermion-antifermion condensate which breaks the global symmetry
to $Sp(2 N_f)$. Since $SU(2N_f)\supset SU_L(N_f)\times
SU_R(N_f)$, the left-right independent groups are unified and
parity invariance is automatic.

This breaking pattern gives $2N^2_f - N_f -1$ Goldstone bosons
which are contained in the antisymmetric matrix $M^{ij}$ and $i,j
=
1,\ldots,2N_f$. With $u\in SU(2N_f)$ we have
\begin{equation}
M\rightarrow u\, M \, u^{T} \ .
\end{equation}
We associate a vector field $A_{\mu}=A_{\mu}^a T^a$ with $T^a $,
a generic generator of $SU(2N_f)$, ($a=1,\ldots, 4N_f^2 -1$) and
$
\displaystyle{{\rm Tr}\left[T^a T^b\right]}=\frac{1}{2}\delta^{ab}$.
Following the procedure outlined in the previous sections, we
define a formal covariant derivative as
\begin{equation}
D_{\mu}M=\partial_{\mu} M - i \tilde{g}A_{\mu} M - i
\tilde{g}MA_{\mu}^{T} \ ,  \label{scd}
\end{equation}
where $A$ transforms as
\begin{equation}
A_{\mu} \rightarrow u A_{\mu} u^{\dagger} - \frac{i}{\tilde{g}}
\partial_{\mu}u u^{\dagger} \ .
\end{equation}

With electroweak interactions turned off, the effective
Lagrangian reads
\begin{eqnarray}
L=&~&\frac{1}{2}{\rm Tr}\left[DM DM^{\dagger}\right] + m^2 \,
{\rm Tr}\left[ A^2\right] + r\,{\rm Tr}\left[A^2 M
M^{\dagger}\right]
\nonumber \\ &+&h\, {\rm Tr}\left[AMA^{T}M^{\dagger}\right] + i\,{s} {\rm
Tr}\left[ A\left(MDM^{\dagger}-DM M^{\dagger}\right)\right] \ ,
\label{spLagrangian}
 \end{eqnarray} together with the kinetic term
\begin{equation}
L_{{\rm Kin}}=-\frac{1}{2}{\rm Tr}\left[F_{\mu \nu}F^{\mu
\nu}\right] \ ,
\end{equation}
where $\displaystyle{F_{\mu \nu}=\partial_{\mu}A_{\nu} -
\partial_{\nu}A_{\mu} } - i \tilde{g}\left[A_{\mu},A_{\nu}\right]$,
 along with globally invariant vector interaction terms, an
interaction term proportional to ${\rm Tr}A^2{\rm Tr}M
M^{\dagger}$, and a symmetry breaking potential.

The global symmetry is enhanced to $SU(2N_f)\times SU(2N_f)$ for
the parameter choice $\displaystyle{s=
\tilde{g},~h=\tilde{g}^2~{\rm and}~r=
\tilde{g}^2}$. The effective Lagrangian then takes the form
\begin{equation}
L=\frac{1}{2}{\rm Tr}\left[\partial M \partial M^{\dagger}\right]
+ m^2
\, {\rm Tr}\left[A^2\right]\,
\end{equation}
together with the same terms as above. The spontaneous breaking
leads to the vacuum symmetry $Sp(2N_f)\times SU(2N_f)$.

To proceed further, we simplify the notation by choosing
$N_{f}=2$. We divide the generators $\{ T
\}$ of $SU(4)$ into two classes, calling the generators of $Sp(4)$
$\{S^{a}\}$ with $a=1,\ldots ,10$ and the broken generators
$\{X^{i}\}$ with $i=1,\ldots ,5$. We have
\begin{equation}
S^{T}\,E+E\,S=0\ ,
\end{equation}
with
\begin{equation}
E=\frac{1}{2\sqrt{2}}\left(
\begin{array}{cc}
{\bf 0} & {\bf 1} \\
-{\bf 1} & {\bf 0}
\end{array}
\right) \ .
\end{equation}
In Appendix~\ref{agenerators} we provide a convenient
representation for the $\{S\}$ and $\{X\}$ generators. We define
the antisymmetric meson matrix $M= (-M^T)$ as
\begin{equation}
M=\sqrt{2}\left[\sigma + i\, 2\sqrt{2} X^i \Pi^i\right]\,E \ ,
\end{equation}
where the five $\Pi^i$ fields are the Goldstone bosons associated
with the breaking of $SU(4)\rightarrow Sp(4)$.

It is convenient to divide the vector field $A$ in the following
way:
\begin{equation} A=A_X + A_S \ ,
\end{equation}
where $A_X=A_X^i\,X^i \ , \quad {\rm and} \quad A_S=A_S^a \, S^a
\
$. The $A_X$ are the axial vector fields while the $A_S$ are the
vectors. Then expanding $M$ around its vacuum value $\sqrt{2}v E$
and keeping only terms quadratic in the fields, the Lagrangian
Eq.~(\ref {spLagrangian}) takes the form
\begin{eqnarray}
L=&~&\frac{1}{2}\left[\partial \sigma \partial \sigma + \partial
\Pi^i
\partial \Pi^i \right]- v\frac{\left(\tilde{g}-s\right)}{\sqrt{2}}
\partial \Pi^i A_X^i + {M^2_X} {\rm Tr} \left[ A_X ^2 \right] + {M^2_S}
{\rm Tr} \left[A_S^2\right],
\end{eqnarray}
with
\begin{eqnarray}
M^2_S&=&m^2 + \frac{v^2}{4} \left[r - h\right] \ ,  \nonumber \\
M^2_X &=& m^2 + \frac{v^2}{4} \left[2\tilde{g}^2 +r -4 \tilde{g}s
+ h \right].
\end{eqnarray}
For the choice of parameters associated with an additional
$SU(4)$ global symmetry (i.e. $\displaystyle{s=\tilde{g},\;
h=\tilde{g}^2}$ and $\displaystyle{r=\tilde{g}^2}$) the
vector-axial vector mass difference vanishes, as do the width to
mass ratios.

We next treat the above theory as an electroweak symmetry
breaking sector by gauging the $SU_{L}(2) \times U_{Y}(1)$
subgroup. It is convenient to introduce a vector field $G_{\mu}$.
If we were to gauge the entire $SU(4)$ flavor symmetry then
$G_{\mu}$ would transform under chiral rotations in the standard
way
\begin{equation}
G_{\mu}\rightarrow u G_{\mu} u^{\dagger} - \frac{i}{g}
\partial_{\mu} u
u^{\dagger} \ .
\end{equation}
We identify the electroweak gauge transformations in the
following way:
\begin{equation} u=\left(
\begin{array}{cc}
u_L & {\bf 0} \\ {\bf 0} & u_R^{\ast}
\end{array}
\right) \ ,
\end{equation}
with $u_{L/R} \in SU_{L/R}(2)$. Then
\begin{equation}
G_{\mu}=\left(
\begin{array}{cc}
W_{\mu} & {\bf 0} \\ {\bf 0} & -\frac{g^{\prime}}{g} B_{\mu}^T
\end{array}
\right) \ ,
\end{equation}
where $\displaystyle{W_{\mu}=W_{\mu}^a \, \frac{\tau^a}{2}}$ and
$
\displaystyle{B_{\mu}=B_{\mu}\, \frac{\tau^3}{2}}$, and $g$ and $g^{\prime}$
are the electroweak couplings. It is easy to verify that the
electroweak transformation properties of the gauge bosons are
respected (see Eq.~(\ref{rotation})).

Using the left-right generators defined in Eq.~(\ref {lrgen}) we
have
\begin{equation} G=W^a L^{a}-\frac{g^{\prime}}{g} B^3 R^{3T} \ ,
\end{equation} with $a=1,2,3$. In terms of the axial and vector
type generators we have \begin{equation} G=G_X + G_S \ ,
\end{equation}
with
\begin{equation}
G_X=\frac{1}{\sqrt{2}}\left(W^a - \frac{g^{\prime}}{g}
B^a\right)X^a\;,
\qquad G_S=\frac{1}{\sqrt{2}}\left(W^a+\frac{g^{\prime}}{g} B^a \right)
S^a \ .
\end{equation}
The covariant derivative including the weak vector bosons and the
composite vector fields is
\begin{equation}
{\rm D}_{\mu}M=\partial_{\mu}M - i\, g \left( G_{\mu} M + M
G_{\mu}^T\right) - i\, \tilde{g}c \left(C_{\mu}M+M C_{\mu}^T
\right) \ ,
\label{CD} \end{equation}
where $c$ is a real coefficient and we have introduced the vector
$C_{\mu}$ \begin{equation} C=A-\frac{g}{\tilde{g}}
\, G \ ,
\end{equation}
transforming covariantly under electroweak rotations.

We extend the effective Lagrangian of Eq.~(\ref{spLagrangian}) to
include electroweak interactions by replacing the old covariant
derivative with the one in Eq.~(\ref{CD}). To render the full
theory invariant under electroweak transformations we also
substitute $A$ with $C$ giving
\begin{eqnarray}
L=&~&\frac{1}{2}{\rm Tr}\left[{\rm D}M {\rm D}M^{\dagger}\right]
+ m^2 \, {\rm Tr}\left[ C^2\right] + r\,{\rm Tr}\left[C^2 M
M^{\dagger}\right]
\nonumber \\ &+&h\, {\rm Tr}\left[CMC^{T}M^{\dagger}\right] + i\,{s} {\rm
Tr}\left[ C\left(M{\rm D}M^{\dagger}-{\rm D}M
M^{\dagger}\right)\right].
\label{sqEW}
\end{eqnarray}
To this we add kinetic terms, interaction terms involving the $C$
fields, the interaction term ${\rm Tr}C^2{\rm Tr}M M^{\dagger}$,
and a symmetry breaking potential.

Replacing $M$ by its vacuum value and retaining only quadratic
mass terms for the vectors we have
\begin{eqnarray}
L=&~&M^2_X (1+\delta) \frac{g^2}{\tilde{g}^2} {\rm
Tr}\left[G^2_X\right] + M^2_X {\rm Tr}\left[A^2_X\right] - 2
\frac{g}{\tilde{g}} M^2_X (1
-\chi){\rm Tr}\left[G_X A_X\right] \nonumber \\ &+& M^2_S {\rm
Tr}\left[A^2_S + \frac{g^2}{\tilde{g}^2}G^2_S - 2\frac{g}{
\tilde{g}}G_S
A_S\right] +\ldots \ ,  \label{spqd} \end{eqnarray} where we have
identified
\begin{eqnarray}
M^2_S&=& m^2 + \frac{v^2}{4}\left[r - h \right]\ ,  \nonumber \\
M^2_X &=& m^2 + \frac{v^2}{4}\left[r + h + 2 \tilde{g}^2 c^2 - 4
s \tilde{g}c
\right]\ ,  \nonumber \\
\delta &=& \frac{v^2}{2M^2_X} \left[\tilde{g}^2 - 2 \tilde{g}\left(\tilde{g}
c - s \right)\right] \ ,  \nonumber \\
\chi &=& \frac{v^2}{2M^2_X} \tilde{g}\left(\tilde{g}c - s \right) \ .
\end{eqnarray}
The generalization of this discussion to the case $N_f > 2$,
necessary for near-criticality of the underlying gauge theory, is
straightforward.

 From this point on, the discussion of enhanced symmetry, parity
degeneracy, and the estimate of the $S$ parameter proceeds as in
the previous section. The choice of parameters $s = \tilde{g}c$
and $h = r = \tilde{g}^{2}c^{2}$ leads to the enhanced symmetry
of the strongly interacting sector and to parity degeneracy. The
contribution to the $S$ parameter from the parity doubled sector
by itself is zero. The contribution from the symmetry breaking
sector is modified by the presence of the larger number of
pseudo-Goldstone bosons~\cite{postmodern} associated with the
$SU(2N_f)$ global symmetry.

\section{Conclusions}
\label{conc}

In this paper we used an effective Lagrangian to explore some
features that might arise in a strongly coupled gauge theory when
the number of fermions $N_f$ is near a critical value for the
transition to chiral symmetry. It has been argued that this
transition is second order or higher and that a long range
conformal symmetry also sets in at the transition. It has also
been suggested that near the transition, parity doublets may
begin to form~\cite{AS,mawhinney}.

We explored this possibility using as a guide an effective
Lagrangian with a linear realization of the global chiral
symmetry. The spectrum was taken to consist of a set of Goldstone
particles, associated massive scalars, and a set of massive
vectors and axial vectors. It was observed that parity doubling
is associated with the appearance of an enhanced global symmetry,
consisting of the spontaneously broken chiral symmetry of the
underlying theory ($SU_{L}(N_f)
\times SU_{R}(N_f)$) together with an additional, unbroken
symmetry, either continuous or discrete. The additional symmetry
leads to the degeneracy of the vector and axial vector, and to
their stability with respect to decay into the Goldstone bosons.

It is worth noting that the effective Lagrangian employed here,
while describing the global symmetries, does not accurately
describe the dynamics of a chiral/conformal transition. That is,
it cannot be used directly as the basis for a Landau-Ginzburg
theory of this transition with its expected nonanalytic behavior
\cite{ATW}. In Ref. \cite{SS}, an approach to such a
Landau-Ginzburg theory was developed, restricted to only the
scalar degrees of freedom, and it described the usual global
symmetries. This approach could perhaps be extended to include
the vectors of the present effective Lagrangian. We expect that
it would describe the same symmetries we have considered here,
both the spontaneously broken symmetry and the additional,
unbroken symmetry.

Despite the hints in Refs. \cite{AS,mawhinney}, it has not been
established that an underlying gauge theory leads to these enhanced
symmetries as $N_f$ approaches a critical value for the chiral
transition. If it is to happen, an unusual and interesting
interplay between confinement and chiral symmetry breaking would
have to develop at the transition.

We also noted, by electroweak gauging of a subgroup of the chiral
flavor group, that the enhanced symmetry provides a partial
custodial symmetry for the $S$ parameter, in that there is no
contribution from the parity-doubled sector by itself. It could
be interesting to explore further the consequences of an enhanced
symmetry for electroweak precision measurements.

\acknowledgments

We thank Sekhar Chivukula for a careful reading of this manuscript.
We also thank Joseph Schechter, Adriano Natale, Noriaki Kitazawa,
Zhiyong Duan, Erich Poppitz, and Alan Chodos for discussions. We
thank Roberto Casalbuoni for helpful comments and for bringing to
our attention some relevant literature. One of us (F.S.) thanks the
theoretical group of Tokyo Metropolitan University for their warm
hospitality. The work of T.A. and F.S. has been partially supported
by the US DOE under contract DE-FG-02-92ER-40704. The work of F.S.
was supported in part by the Grant-in-Aid for Scientific Research
\# 0945036 under the International Scientific Research Program,
Inter- University Cooperative Research. The work of P.S.R.S. has
been supported by the Funda\c{c}\~ao de Amparo \`a Pesquisa do
Estado de S\~ao Paulo (FAPESP) under contract \# 98/05643-5.

\appendix

\section{Explicit Realization of the $Sp(4)$ Generators}

\label{agenerators}

We conveniently represent the generators of $SU(4)$ in the
following way
\begin{equation}
S^{a}=\left(
\begin{array}{cc}
{\bf A} & {\bf B} \\ {\bf B^{\dagger }} & -{\bf A}^{T}
\end{array}
\right) \ ,\qquad X^{i}=\left(
\begin{array}{cc}
{\bf C} & {\bf D} \\ {\bf D^{\dagger }} & {\bf C}^{T} \end{array}
\right) \ ,
\end{equation}
where ${A}$ is hermitian, ${C}$ is hermitian and traceless,
${B}={B}^{T}$ and ${D}=-{D}^{T}$. The $\{S\}$ are also a
representation of the $Sp(4)$ generators since they obey the
relation $S^{T}E+ES=0$. We define
\begin{equation} S^{a}=\frac{1}{2\sqrt{2}}\left( \begin{array}{cc}
\tau ^{a} & {\bf 0} \\
{\bf 0} & -\tau ^{aT}
\end{array}
\right) \ ,\qquad a=1,2,3,4\ .
\end{equation}
{}For $a=1,2,3$ we have the standard Pauli matrices, while for
$a=4$ we define $\tau ^{4}={\bf 1}$. These are the generators for
$SU_V(2)\times U_V(1)$. {}For $a=5,\ldots,10$
\begin{equation}
S^{a}=\frac{1}{2\sqrt{2}}\left(
\begin{array}{cc}
{\bf 0} & {\bf B}^{a} \\ {\bf B}^{{a\dagger }} & {\bf 0}
\end{array}
\right) \ , \qquad a=5,\dots,10
\end{equation}
and
\begin{equation}
\begin{array}{ccc}
B^5=1 & B^7=\tau^3 & B^9 = \tau^1 \\ B^6 = i\, 1 & B^8=i\, \tau^3
& B^{10} = i \tau^1 \end{array}
\end{equation}
The five axial type generators $\{X^i\}$ are \begin{equation}
X^{i}=\frac{1}{2\sqrt{2}}\left(
\begin{array}{cc}
\tau ^{i} & {\bf 0} \\
{\bf 0} & \tau ^{iT}
\end{array}
\right) \ ,\qquad i=1,2,3\ .
\end{equation}
$\tau^i$ are the standard Pauli matrices. {}For $i=4,5$
\begin{equation}
X^{i}=\frac{1}{2\sqrt{2}}\left(
\begin{array}{cc}
{\bf 0} & {\bf D}^{i} \\ {\bf D}^{{i\dagger }} & {\bf 0}
\end{array}
\right) \ , \qquad i=4,5 \ ,
\end{equation}
and
\begin{equation}
D^4=\tau^2 \ , \qquad D^5=i\,\tau^2 \ .
\end{equation}
The generators are normalized as follows \begin{equation} {\rm
Tr}\left[ S^{a}S^{b}\right] ={\rm Tr}\left[ X^{a}X^{b}\right]
=\frac{1}{2}\delta
^{ab}\ ,\qquad {\rm Tr}\left[ X^{i}S^{a}\right] =0\ .
\end{equation}

The $SU_{L/R}(2)$ generators are readily identified as
\begin{eqnarray}
L^a &\equiv &\frac{S^a + X^a}{\sqrt{2}} \ , \\ R^a &\equiv
&\frac{X^{aT}- S^{aT}}{\sqrt{2}} \ ,  \label{lrgen}
\end{eqnarray} and $a=1,2,3$.

\end{document}